\documentclass[11pt]{cernrep}
\usepackage{graphicx}

\begin{document}
\title{REGULARIZATION OF A LIGHT-FRONT Qqq MODEL
\footnote{{\bf \ \ Talk given in "Light-Cone Workshop:
Hadrons and Beyond", \ Durham, \ 2003.}}}
\author{E. F. Suisso $^{a}$, J.P.B.C. de Melo $^{b}$, T. Frederico $^{a}$}
\institute{$^{a}$ Dept. de F\'{i}sica, Instituto Tecnol\'{o}gico
da Aeron\'{a}utica, Centro T\'{e}cnico Aeroespacial, 12228-901,
S\~{a}o Jos\'{e} dos Campos, SP, Brazil. $^{b}$ Instituto de
F\'{i}sica Te\'{o}rica, Universidade Estadual Paulista, 01450-900,
S\~{a}o Paulo, SP, Brazil} \maketitle
\begin{abstract}
We study the mass of the ground state of $Qqq$ systems using
different regularization schemes of the relativistic integral
equation obtained with a flavor independent contact interaction in
a QCD-inspired light-front model. We calculate the masses of the
spin 1/2 low-lying states of the $\Lambda ^{0}$, $\Lambda
_{c}^{+}$ and $\Lambda _{b}^{0}$ for different values of the
regularization cut-off parameter with a fixed nucleon mass. Our
results are in remarkable agreement with the experimental data.
\end{abstract}

\section{INTRODUCTION}

The QCD-inspired light-front constituent quark
model~\cite{pauli,brodsky}, with two components in the
interaction, a contact term and a Coulomb-like potential, was used
previously to investigate the properties of
mesons~\cite{pauli,tobpauli} with reasonable success. It was also
applied to investigate the binding energy of the ground state of
spin 1/2 $Qqq$ baryons~\cite{Sui02}, where the Coulomb-like
interaction was left out. In that work, a special regularization
scheme was used in which the masses of the virtual two-body
subsystems were constrained to be real, as a result the quark
binding in the spin 1/2 low-lying states of the $\Lambda ^{0}$,
$\Lambda _{c}^{+}$ and $\Lambda _{b}^{0}$ was qualitatively
reproduced. Recently, the integral equation for a three-boson
system interacting with pairwise contact interaction in the
light-front \cite{Fred94} was regularized with a sharp cut-off
\cite{beyer03} and applied in the study of the nucleon. In the
infinite cut-off limit it was previously shown that the
three-boson system is stable for values of the two-boson bound
state mass above a critical value \cite{karmanov02}. This motivate
us to study the effect of different cut-offs in the $Qqq$ ground
state masses, obtained by solving the light-front integral
equations with a contact interaction. Our aim here is to check to
which extend the qualitative properties of the quark binding is
regularization independent.

We use only the flavor independent contact interaction between the
constituent quarks, which brings the physical scale of the ground
state of the nucleon and includes the minimal number of physical
scales to describe $Qqq$ systems. The spin is averaged out. The
model has as inputs the constituent quark masses and the nucleon
mass is used to fix the interaction strength. In our previous work
\cite{Sui02}, where the masses of the virtual two-quark subsystems
were constrained to be real with the nucleon mass fixed at the
experimental value, there was no freedom left in the $Qqq$
calculation. Here, even with the nucleon mass kept fixed still the
cut-off has some freedom, which will be varied in our
calculations. To obtain the $Qqq$ masses, the mass of one of the
constituent quark $(Q)$ is varied while the strength of the
effective contact interaction is supposed flavor independent.

This work is organized as follows. In Sec.2, we briefly discuss
the coupled integral equations for the $Qqq$ bound-state in the
light-front and the regularization schemes. In Sec. 3, we present
the numerical results for the binding energies of the $\Lambda
^{0}$, $\Lambda _{c}^{+}$ and $\Lambda _{b}^{+}$ and give a
summary of the work with our conclusion.

\section{INTEGRAL EQUATIONS FOR THE $Qqq$ BOUND STATE  }

With the assumption that the spin is averaged out, the $Qqq$
system interacting through a contact pairwise force is represented
by two spectator functions or Faddeev components of the vertex,
which satisfies the coupled  Bethe-Salpeter type equations,
derived elsewhere \cite{Sui02}, and shown diagrammatically in
Fig.~1.

\begin{figure}
\begin{center}
\includegraphics[width=7cm]{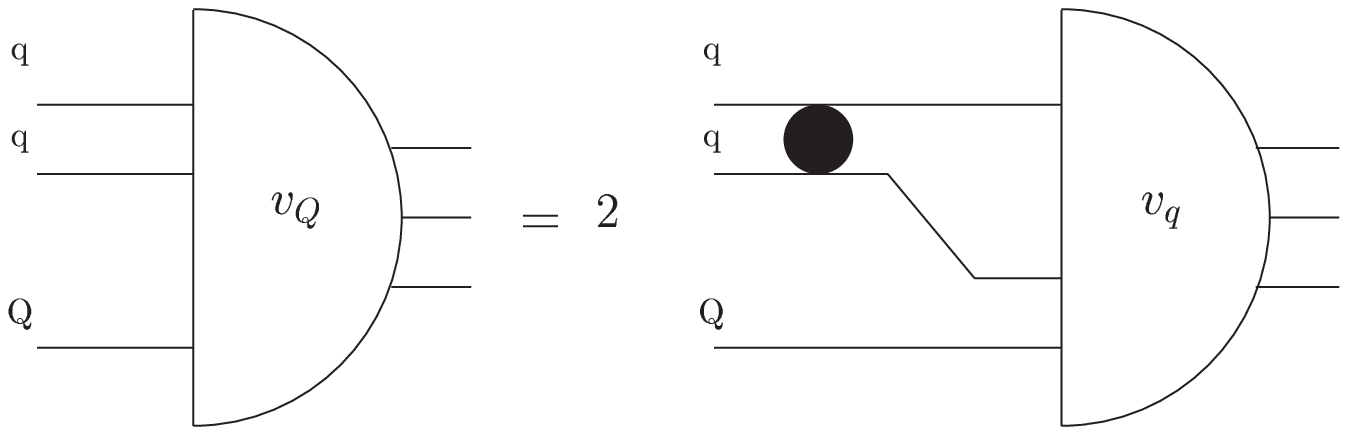}
\includegraphics[width=9cm]{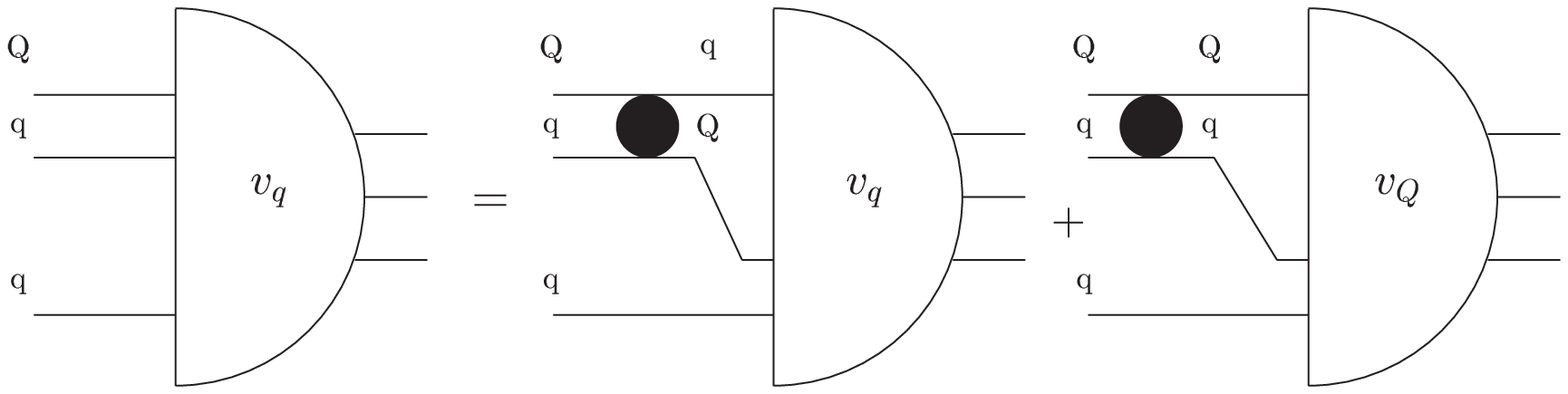}
\caption{Diagrammatic representation of the coupled light-front
Bethe-Salpeter equations of the $Qqq$ system.  }
\end{center}
\end{figure}

The coupled integral equations for the Fadeev components of the
vertex function of the $Qqq$ system,  represented in Fig.~1, are
given by~\cite{Sui02}:

\begin{eqnarray}
v_{Q}\left( \overrightarrow{q}_{\bot },y\right) &=&\frac{i}{\left(
2\pi \right) ^{3}}\tau _{\left( qq\right) }\left(
M_{qq}^{2}\right)
\int\limits_{0}^{1-y}\frac{dx}{x\left( 1-x-y\right) }\times  \nonumber \\
&&\int\limits_{}^{}d^{2}k_{\bot }\frac{\theta \left( x-\frac{%
m_{q}^{2}}{M_{3B}^{2}}\right) \theta \left( k^{\max }_\bot-k_{\bot }\right) }{%
M_{3B}^{2}-\frac{q_{\bot }^{2}+m_{Q}^{2}}{y}-\frac{k_{\bot }^{2}+m_{q}^{2}}{x%
}-\frac{\left( P_{3B}-q-k\right) _{\bot }^{2}+m_{q}^{2}}{x\left(
1-x-y\right) }}v_{q}\left( \overrightarrow{k}_{\bot },x\right) \ ,
\label{Mass3}
\end{eqnarray}

\begin{eqnarray}
v_{q}\left( \overrightarrow{q}_{\bot },y\ \right)
&=&2\frac{i}{\left( 2\pi \right) ^{3}}\tau _{\left( Qq\right)
}\left( M_{Qq}^{2}\right)
\int\limits_{0}^{1-y}\frac{dx}{x\left( 1-x-y\right) }\times   \nonumber \\
&&\left[ \int\limits_{}^{}d^{2}k_{\bot }\frac{\theta \left( x-%
\frac{m_{Q}^{2}}{M_{3B}^{2}}\right) \theta \left( k^{\max
}_\bot-k_{\perp }\right) }{M_{3B}^{2}-\frac{q_{\bot
}^{2}+m_{q}^{2}}{y}-\frac{k_{\bot
}^{2}+m_{Q}^{2}}{x}-\frac{\left( P_{3B}-q-k\right) _{\bot }^{2}+m_{q}^{2}}{%
\left( 1-x-y\right) }}v_{Q}\left( \overrightarrow{k}_\bot,x\right)
\right.   \nonumber \\
&&\left. +\int\limits_{}^{}d^{2}k_{\bot }\frac{\theta \left( x-%
\frac{m_{q}^{2}}{M_{3B}^{2}}\right) \theta \left( k^{\max
}_\bot-k_{\perp }\right) }{M_{3B}^{2}-\frac{q_{\bot
}^{2}+m_{q}^{2}}{y}-\frac{k_{\bot
}^{2}+m_{q}^{2}}{x}-\frac{\left( P_{3B}-q-k\right) _{\bot }^{2}+m_{Q}^{2}}{%
\left( 1-x-y\right) }}v_{q}\left( \overrightarrow{k}_{\bot
},x\right) \right] ,  \label{E26}
\end{eqnarray}
where $M_{3B}$ and $P_{3B}$ are the $Qqq$ mass and four-momentum,
respectively. The masses of the virtual two-quark subsystems are
$M_{qq}$ and $M_{Qq}$, and in the baryon rest frame are given by:

\begin{equation}
M_{\alpha q}^{2}=\left( P_{3B}-q\right) ^{2}= \left( 1-y\right) \left( M^2_{3B}-%
\frac{q_{\perp }^{2}+m_{X}^{2}}{y}\right) -q_{\perp }^{2} \ ,
\label{Mass3a}
\end{equation}
with ($\alpha$,$X$)=$(q,Q)$ or $(Q,q)$. In our previous
work~\cite{Sui02}, we constrained the  virtual two-body subsystem
masses to real values and using Eq.~(\ref{Mass3a}) with $q_\bot$
and $y$ substituted by $k_\bot $ and $x$, respectively, giving the
lower bounds for the integration in $x$ and the maximum value for
the transverse momentum, $k^{max}_\bot$, for each vertex depending
on $x$, as well. In that case, the theta functions bring to
Eqs.~(\ref{Mass3}) and (\ref{E26}) the discussed constraint. The
two-quark scattering amplitudes, $\tau _{\alpha q}\left( M_{\alpha
q}^{2}\right) $,  are the solutions of the Bethe-Salpeter
equations in the ladder approximation for a contact interaction
between the quarks.

\begin{figure}%[thbp]
\begin{center}
\includegraphics[width=8.5cm]{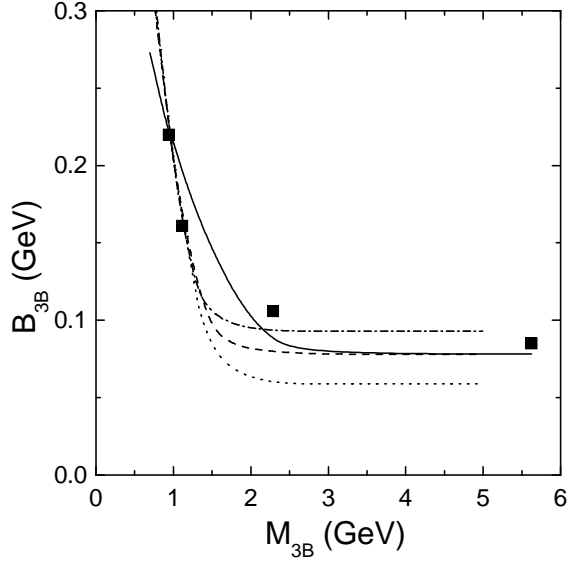}
\caption{Binding energy $B_{3B}$ as a function of the baryon mass
$M_{3B}$. Attributed experimental binding energies for the spin
1/2 low-lying states of the nucleon, $\Lambda ^{0}$, $\Lambda
_{c}^{+}$ and $\Lambda _{b}^{0}$ from \cite{Sui02} (full squares).
Model results from ref. \cite{Sui02} (solid curve). Present
calculations with transverse momentum cut-off with the values of
$3m_q$ (dashed curve), $5m_q$ (dotted curve) and $7m_q$
(dot-dashed curve).}
\end{center}
\end{figure}

Here, we use a different regularization scheme of Eqs.
(\ref{Mass3}) and (\ref{E26}), where  $k^{max}_\bot=\Lambda_\perp$
and the theta functions in $x$ are dropped. The nucleon mass is
fixed to its experimental value, therefore we have to adjust the
renormalized coupling constant for each value of the transverse
momentum cut-off

\begin{eqnarray}
i~\tau _{\alpha q}\left( M_{\alpha q}^{2}\right) =\left[ \lambda
^{-1}- \frac{1}{2\left( 2\pi \right)
^{3}}\int\limits_{0}^{1}\frac{dxd^{2}p_{\bot }}{x\left( 1-x\right) }\frac{1}{%
M_{\alpha q}^{2}-\frac{p_{\bot }^{2}+\left( m_{\alpha
}^{2}-m_{q}^{2}\right) x+m_{q}^{2}}{x\left( 1-x\right)
}}\right]^{-1}\ , \label{tau}
\end{eqnarray}
where $\lambda $ is the bare interaction strength. As its stands,
Eq.~(\ref{tau}) is ill-defined. The renormalization condition is
chosen at an arbitrary subtraction mass point $ \mu $ where the
value of $\tau _{q q}\left( \mu^{2}\right)$ is known and given by
$\lambda _{ren}^{-1}\left( \mu^{2}\right)$, the renormalized
interaction strength, which is enough  to remove the logarithmic
divergence in Eq.~(\ref{tau}). Therefore, the bare strength is
given as:

\begin{equation}
\lambda ^{-1} = \lambda _{ren}^{-1}\left(
\mu^{2}\right)+\frac{1}{2\left( 2\pi \right)
^{3}}\int\limits_{0}^{1}\frac{dxd^{2}p_{\bot }}{x\left( 1-x\right) }\frac{1}{%
\mu^{2}-\frac{p_{\bot }^{2}+  m_{q }^{2} }{x\left( 1-x\right) }},
\end{equation}
which is supposed to be flavor independent. The strength of the
effective interaction is determined by fitting the nucleon mass.

\section{RESULTS AND SUMMARY}

The physical inputs of the model defined by Eqs. (\ref{Mass3}) and
(\ref{E26}), are the renormalized interaction strength
$(\lambda_{ren}(\mu^2))$, the constituent quark masses
$m_q=m_{u,d}=0.386$ GeV   and  $m_{Q}$ ($ Q=s,c,b$) which were
found in \cite{Sui02}, where as well, it was attributed a binding
energy to  the spin 1/2 baryons (nucleon, $\Lambda ^{0}$, $\Lambda _{c}^{+}$ and $%
\Lambda _{b}^{0})$ defined as $B_{3B}=2m_{q}+m_{Q}-M_{3B}$.
Considering that the flavor-off-diagonal vector mesons are weakly
bound systems of constituent quarks  in the QCD-inspired model of
Ref.~\cite{tobpauli}, we obtained values of the constituent quark
masses. The attributed values to the  binding energies were
found~\cite{Sui02} using constituent quark masses and the
experimental baryon masses from \cite{pdg}. The attributed values
for the constituent quark binding energies in the spin 1/2 low-lying states of the nucleon, $%
\Lambda ^{0}$, $\Lambda _{c}^{+}$ and $\Lambda _{b}^{0}$ are shown
in Fig.~2 by the full squares, from the left to the right,
respectively.

In Fig.~2, we show results for the calculation of the binding
energy of the baryon for different renormalization schemes and
values of the transverse momentum cut-off compared to the
attributed baryon binding energies. For  the values of the
cut-off, $3m_q$, $5m_q$ and $7m_q$, we fit the renormalized
strength of the interaction to reproduce the nucleon mass. By
changing the mass of the heavy quark, while the strength of the
interaction was kept unchanged, the coupled integral equations
were solved to obtain the mass and binding energy of the baryons.
Again, as observed in the work \cite{Sui02} for the baryon mass
above $2.3$ GeV, the bound $Qqq$ system goes to the diquark
threshold.  Our results show that the general behavior of the
experimental data of masses and binding energies of the spin 1/2
low-lying states of $\Lambda ^{0} $, $\Lambda _{c}^{+}$ and
$\Lambda _{b}^{0}$ are reproduced with different momentum
cut-off's constrained by the value of the nucleon mass.

In summary, in the present contribution we have studied how
different regularization schemes in the relativistic coupled $Qqq$
homogeneous integral equation affects the qualitative properties
of the quark binding in $Qqq$ baryons. Our calculations indicate
that the general features of the low-lying spin 1/2 baryon binding
energy as a function of its mass, is model independent to a large
extent (small cut-off sensitivity) as long as the relativistic
$Qqq$ model has a flavor independent effective short-range
interaction. Therefore, the conclusions drawn in our previous work
\cite{Sui02} does not qualitatively depend on the regularization
scheme.

\noindent

\section*{ACKNOWLEDGEMENTS}

We would like to thank the Brazilian funding agencies FAPESP
(Funda\c{c}\~ao de Amparo a Pesquisa do Estado de S\~ao Paulo) and
CNPq (Conselho Nacional de Desenvolvimento Cient\'{\i}fico e
Tecnol\'ogico). EFS thanks the Organizing Committee of the
conference for the kindly invitation to present this work.

\end{document}